\documentclass[]{mn2e}
\usepackage{graphicx}
\title[Monthly Notices: \LaTeXe\ guide for authors]
  {Support Vector Machines and Kd-tree for Separating Quasars from Large Survey Databases}
\author[D. Gao et al.]
  {Dan Gao$^{1,2}$
  ,Yan-Xia Zhang$^1$\thanks{Email:zyx@lamost.org}
  and Yong-Heng Zhao$^1$ \\
  $^1$National Astronomical Observatories,
Chinese Academy of Sciences, 20A Datun Road, Chaoyang District,
100012, Beijing, P.R.China\\
  $^2$Graduate University of Chinese Academy of Sciences,
Beijing, P.R.China}
\date{Released 2002 Xxxxx XX}

\pagerange{\pageref{firstpage}--\pageref{lastpage}} \pubyear{2002}

\def\LaTeX{L\kern-.36em\raise.3ex\hbox{a}\kern-.15em
    T\kern-.1667em\lower.7ex\hbox{E}\kern-.125emX}

\begin{document}

\label{firstpage}

\maketitle

\begin{abstract}
We compare the performance of two automated classification
algorithms: k-dimensional tree (kd-tree) and support vector machines
(SVMs), to separate quasars from stars in the databases of the Sloan
Digital Sky Survey (SDSS) and the Two Micron All Sky Survey (2MASS)
catalogs. The two algorithms are trained on subsets of SDSS and
2MASS objects whose nature is known via spectroscopy. We choose
different attribute combination as input patterns to train the
classifier using photometric data only and present the
classification results obtained by these two methods. Performance
metrics such as precision and recall, true positive rate and true
negative rate, F-measure, G-mean and Weighted Accuracy are computed
to evaluate the performance of the two algorithms. The study shows
that both kd-tree and SVMs are effective automated algorithms to
classify point sources. SVMs show slightly higher accuracy, but
kd-tree requires less computation time. Given different input
patterns based on various parameters(e.g. magnitudes, color
information), we conclude that both kd-tree and SVMs show better
performance with fewer features. What is more, our results also
indicate that the accuracy using the four colors ($u-g$, $g-r$,
$r-i$, $i-z$) and $r$ magnitude based on SDSS model magnitudes adds
up to the highest value. The classifiers trained by kd-tree and SVMs
can be used to solve the automated classification problems faced by
the virtual observatory (VO); moreover, they all can be applied for
the photometric preselection of quasar candidates for large survey
projects in order to optimize the efficiency of telescopes.

\end{abstract}

\begin{keywords}
 Classification, Astronomical databases:
miscellaneous, Catalogs, Methods: Data Analysis, Methods:
Statistical
\end{keywords}

\section{Introduction}
In the recent years, the sizes of astronomical data based on surveys
at different wavebands are increasing rapidly. Astronomy has entered
a data avalanche era. The most important and challenging issues for
the efficient analysis of large multi-wavelength astronomical data
rely on data mining tools, which will allow the selection,
classification, regression, clustering and even the definition of
particular object types within the databases.

Our primary goal is to perform reliable star-quasar separation.
Since stars and quasars are point sources, their classification is
an important issue in astronomy. In the recent past a lot of work
has been carried out on automated approaches. Hatziminaoglou et al.
(2000) explored a new joint method (avoiding usual biases) for
distinguishing between quasars and stars/galaxies by their
photometry. Wolf et al. (2004) explored a photometric method for
identifying stars, galaxies and quasars in multi-color surveys.
McGlynn (2004) used decision trees to build an online system for
automated classification of X-ray sources. Carballo et al. (2004)
selected quasar candidates from combined radio and optical surveys
using neural networks. Suchkov et al. (2005) applied ClassX, an
oblique decision tree classifier optimized for astronomical
classification and redshift estimation in the Sloan Digital Sky
Survey (SDSS) photometric catalog. Ball et al. (2006) classified
stars and galaxies with the SDSS DR3 using decision trees.

In this work we investigate the application of support vector
machines (SVMs) and k-dimensional tree (kd-tree) to effectively
select quasar candidates. SVMs have been successfully applied in
astronomy for mainly the following problems: classification of
variable stars (Wozniak et~al. 2001, 2004), galaxy morphology
classification (Humphreys et~al. 2001), solar-flare detection (Qu
et~al. 2003), classification of multiwavelength data (Zhang \& Zhao
2003, 2004), estimation of photometric redshifts of galaxies
(Wadadekar 2005; Wang et~al. 2007) and matching different object
catalogs in astrophysics (Rohde et~al. 2005, 2006). Wang et~al.
(2007) investigated SVMs and Kernel Regression (KR) for photometric
redshift estimation with the data from the SDSS Data Release 5 (DR5)
and the Two Micron All Sky Survey (2MASS). On the other hand, the
kd-tree method is used on the 5 flux-space indexing in the SDSS
science archive to partition the bulk data (Kunszt et~al. 2000).
Maneewongvatana \& Mount (2002) presented an empirical analysis of
two new splitting methods for kd-trees: sliding-midpoint and
minimum-ambiguity, which were designed to remedy some of the
deficiencies of the standard kd-tree splitting method, with respect
to data distributions that are highly clustered in low-dimensional
subspaces. Hsieh et~al. (2005) used kd-tree algorithm to divide
their sample in order to improve the redshift accuracy of galaxies.
Kubica et~al. (2007) employed kd-tree for efficient intra- and
inter-night linking of asteroid detections. Gao et~al. (2008)
introduced some application cases of kd-tree in astronomy. In real
application, estimation of photometric redshifts belongs to
regression problem. Although SVMs and kd-tree are applied for both
classification and regression problems, they can't solve them
simultaneously. For classification problem, the predicted parameter
is discrete; while for regression problem, the predicted parameter
is continuous. When dealing with the two different tasks, the
methods should be adjusted.

The structure of this paper is as follows: Section~2 gives the
sample collection and parameter selection. Section~3 presents the
brief introduction of kd-tree and SVMs. Section~4 illustrates the
results and discussion, and the conclusion is presented in
Section~5.

\section{Sample and Parameter Selection}

The Sloan Digital Sky Survey (SDSS, York et~al. 2000) uses a
dedicated, wide field, 2.5m telescope at Apache Point Observatory,
New Mexico. Imaging is carried out in drift-scan mode using a 142
mega-pixel camera in five broad bands, $u$ $g$ $r$ $i$ $z$, spanning
the range from 3000 to 10,000$\AA$. The corresponding magnitude
limits for the five bands are 22.0, 22.2, 22.2, 21.3 and 20.5,
respectively. The Fifth Data Release (DR5) of the SDSS includes all
survey quality data taken through June 2005 and represents the
completion of the SDSS-I project. It includes five-band photometric
data for 215 million unique objects selected over 8000~deg$^2$ , and
1,048,960 spectra of galaxies, quasars, and stars selected from 5740
deg$^2$ of that imaging data. The magnitude limits for the
spectroscopic samples are $r(Petrosian)$=17.77 for the galaxies and
$i(PSF)$=19.1 for quasars with redshifts up to 2.3 and $i(PSF)$=20.1
for quasars with higher redshifts.

The Two Micron All Sky Survey (2MASS) project (Cutri et~al. 2003) is
designed to close the gap between our current technical capability
and our knowledge of the near-infrared sky. 2MASS uses two new,
highly-automated 1.3m telescopes, one at Mt. Hopkins, AZ, and one at
CTIO, Chile. Each telescope is equipped with three-channel camera,
each channel consisting of 256x256 array of HgCdTe detectors,
capable of observing the sky simultaneously at $J$ (1.25$\mu$m), $H$
(1.65$\mu$m) and $K_s$ (2.17$\mu$m), to 3$\sigma$ limiting
sensitivity of 17.1, 16.4 and 15.3 mag in the three bands. The
number of 2MASS point sources adds up to 470,992,970.

We collected photometric data of quasars and stars with spectra
measurement from SDSS DR5, then cross-identified the 2MASS database
with these photometric data within a 2~arcsec radius by the
federation system of XMaS\_VO. XMaS\_VO is developed by China-VO
project and mainly used for automation of creating databases and
cross-identification of catalogues from different bands (Gao et~al.
2008). We obtained the samples, as shown in Table~1. The result
shows that almost every SDSS object has the counterpart in the 2MASS
database and there are only less than 100 missing data records. In
our work, for all objects under consideration the SDSS and 2MASS
magnitudes are available. In this way the issue of inhomogeneous
coverage or non-detections is not dealt with.

\begin{table*}
 \caption{The number of samples from different catalogs }
 \label{symbols}
 \begin{tabular}{@{}lcccccc}
  \hline
  Catalog & Number of Quasars & Number of Stars\\
  \hline
SDSS&76,949&108,744\\
SDSS+2MASS&76,863&108,679\\

   \hline
 \end{tabular}

 \end{table*}

In order to study the distribution of stars and quasars in the
multi-dimensional space, we use different magnitudes: PSF magnitude
($u^p$ $g^p$ $r^p$ $i^p$ $z^p$), model magnitude ($u$ $g$ $r$ $i$
$z$) and model magnitude with reddening correction ($u'$ $g'$ $r'$
$i'$ $z'$, hereafter short for dereddened magnitude) from SDSS data,
$J$, $H$ and $K_s$ magnitudes from 2MASS catalog. The dereddened
magnitudes are corrected by Galaxy extinction using the dust maps of
Schlegel et~al. 1998.  $J$, $H$ and $K_s$ is the selected ``default"
magnitude for each band, respectively. If the source is not detected
in the band, this is the 95\% confidence upper limit derived from a
4 arcsec radius aperture measurement taken at the position of the
source on the Atlas Image. We explored different input patterns
composed of these attributes.

The mean values of the features selected as input patterns are given
in Table~2, which shows the statistical properties of the samples.
The first, second and third columns give the number, name and
description of the parameters, respectively. The following columns
list the mean values of the parameters with standard errors for
quasars and stars. Obviously, the mean values of parameters of the
samples are different for different classes of objects, especially
for the color indexes. Therefore it is reasonable and applicable to
discriminate quasars from stars with these features. In order to
investigate the distribution of different objects in 2D scatter
plots, we randomly select some parameters and subsamples of quasars
and stars for visual inspection and plot them in Figure 1.

Taking the pattern ($u-g$, $g-r$, $r-i$, $i-z$, $r$) for example, we
apply principal component analysis (PCA) on the sample. PCA is a
statistical method that permits the determination of the minimum
number of independent or uncorrelated variables underlying a larger
number of observed variables (Kendall 1957; Kendall \& Stuart 1966).
Thus, PCA is used as a technique for both data compression and
analysis, in addition, PCA can be used as an unsupervised method for
classification. As for PCA used in astronomy, we refer to e.g.
Connolly \& Szalay (1999 and references therein) or Zhang \& Zhao
(2003). The result of PCA shows that the first three eigenvectors
carry 99.30\%, 0.41\% and 0.17\%, respectively, of the descriptive
power. This means that the first three vectors actually carry most
of the information, especially the first one. We study the
distribution of quasars and stars in the principal component space.
To be simple, the subsample randomly selected from the overall
sample is shown in Figure 1. PC1, PC2 and PC3 are short for the
first, second and third principal components, respectively.

It is obvious from Figure 1 that quasars and stars are not easy to
discriminate from each other due to overlapping in the two-color
diagrams or the principal component spaces. Therefore we need rely
on machine learning or data mining techniques to realize the
separation between quasars and stars in the high dimensional space .

\begin{table*}
 \caption{The mean values of parameters for the samples. }
 \label{symbols}
 \begin{tabular}{@{}lcccccc}
  \hline
  No. & Parameters& Desciption & Quasars
        & Stars \\
  \hline
  1 ...... & $u^p$ &SDSS PSF $u$ magnitude& $19.78\pm1.38$ & $20.78\pm2.57$ \\

  2 ...... & $g^p$ &SDSS PSF $g$ magnitude& $19.31\pm0.89$ & $19.32\pm2.23$  \\

  3 ...... & $r^p$ &SDSS PSF $r$ magnitude& $19.06\pm0.76$ & $18.56\pm1.81$  \\

  4 ...... & $i^p$ &SDSS PSF $i$ magnitude& $18.89\pm0.75$ & $18.02\pm1.51$  \\

  5 ...... & $z^p$ &SDSS PSF $z$ magnitude& $18.80\pm0.76$ & $17.73\pm1.48$  \\

  6 ...... & $u^p-g^p$ &SDSS PSF $u-g$ color& $0.48\pm0.79$ & $1.46\pm1.04$ \\

  7 ...... & $g^p-r^p$ &SDSS PSF $g-r$ color& $0.25\pm0.35$ & $0.76\pm0.81$ \\

  8 ...... & $r^p-i^p$ &SDSS PSF $r-i$ color& $0.16\pm0.20$ & $0.54\pm0.87$  \\

  9 ...... & $i^p-z^p$ &SDSS PSF $i-z$ color& $0.10\pm0.18$ & $0.30\pm0.59$  \\

  10...... & $u$ &SDSS model $u$ magnitude& $19.74\pm1.41$ & $20.77\pm2.63$  \\

  11...... & $g$ &SDSS model $g$ magnitude& $19.23\pm0.93$ & $19.26\pm2.24$  \\

  12...... & $r$ &SDSS model $r$ magnitude& $18.96\pm0.84$ & $18.49\pm1.83$  \\

  13...... & $i$ &SDSS model $i$ magnitude& $18.81\pm0.85$ & $17.96\pm1.52$  \\

  14...... & $z$ &SDSS model $z$ magnitude& $18.71\pm0.87$ & $17.67\pm1.50$  \\

  15...... & $u-g$ &SDSS model $u-g$ color& $0.50\pm0.81$ & $1.50\pm1.09$  \\

  16...... & $g-r$ &SDSS model $g-r$ color& $0.26\pm0.37$ & $0.78\pm0.85$  \\

  17...... & $r-i$ &SDSS model $r-i$ color& $0.17\pm0.21$ & $0.53\pm0.89$  \\

  18...... & $i-z$ &SDSS model $i-z$ color& $0.10\pm0.18$ & $0.29\pm0.62$  \\

  19...... & $u'$ &SDSS dereddened model $u$ magnitude& $19.58\pm1.41$ & $20.58\pm2.63$  \\

  20...... & $g'$ &SDSS dereddened model $g$ magnitude& $19.12\pm0.93$ & $19.13\pm2.24$  \\

  21...... & $r'$ &SDSS dereddened model $r$ magnitude& $18.89\pm0.84$ & $18.39\pm1.83$  \\

  22...... & $i'$ &SDSS dereddened model $i$ magnitude& $18.74\pm0.85$ & $17.88\pm1.52$  \\

  23...... & $z'$ &SDSS dereddened model $z$ magnitude& $18.87\pm0.87$ & $17.61\pm1.50$  \\

  24...... & $u'-g'$ &SDSS dereddened model $u-g$ color& $0.46\pm0.81$ & $1.45\pm1.09$  \\

  25...... & $g'-r'$ &SDSS dereddened model $g-r$ color& $0.23\pm0.37$ & $0.74\pm0.85$  \\

  26...... & $r'-i'$ &SDSS dereddened model $r-i$ color& $0.15\pm0.21$ & $0.51\pm0.89$  \\

  27...... & $i'-z'$ &SDSS dereddened model $i-g$ color& $0.08\pm0.18$ & $0.27\pm0.62$  \\

  28...... & $J$ &2MASS $J$ magnitude& $15.58\pm1.39$ & $15.39\pm1.21$  \\

  29...... & $H$ &2MASS $H$ magnitude& $15.00\pm1.33$ & $14.91\pm1.21$  \\

  30...... & $K_s$ &2MASS $K_s$ magnitude& $14.67\pm1.23$ & $14.72\pm1.19$  \\

  31...... & $J-H$ &2MASS $J-H$ color& $0.59\pm0.27$ & $0.48\pm0.25$  \\

  32...... & $H-K_s$ &2MASS $H-K_s$ color& $0.32\pm0.37$ & $0.19\pm0.31$  \\

  \hline
 \end{tabular}

 \medskip
\end{table*}

\begin{figure*}
   \begin{center}

\includegraphics[bb=0 0 300 240,width=6cm,clip]{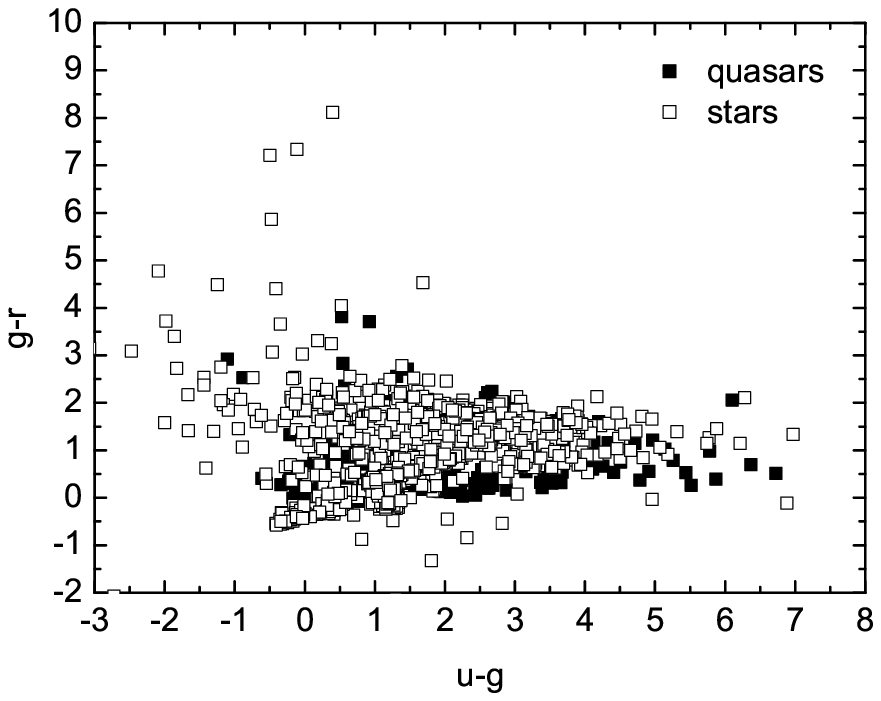}
\includegraphics[bb=0 0 300 240,width=6cm,clip]{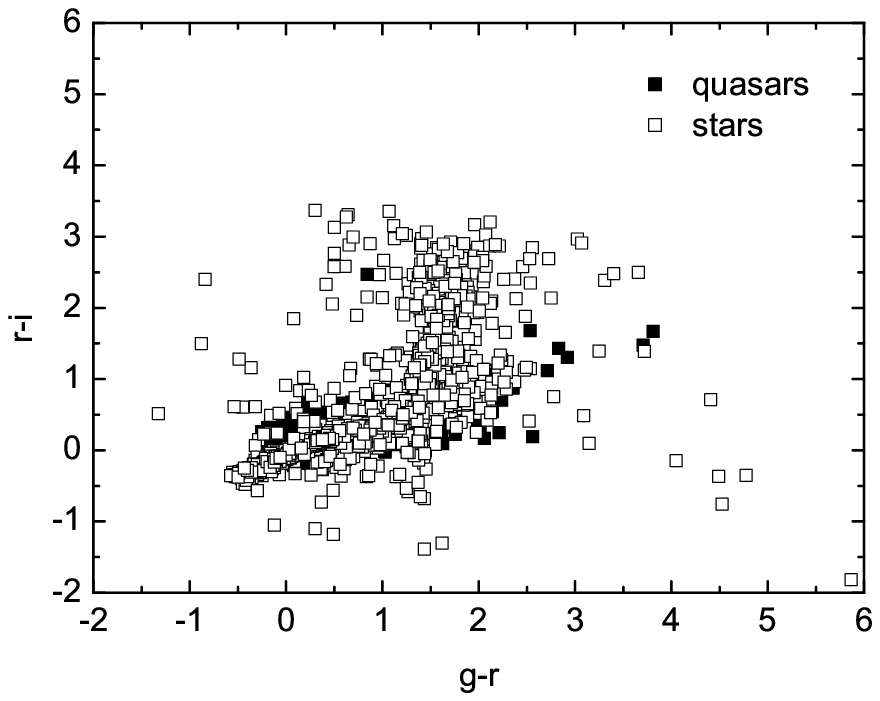}
\includegraphics[bb=0 0 300 240,width=6cm,clip]{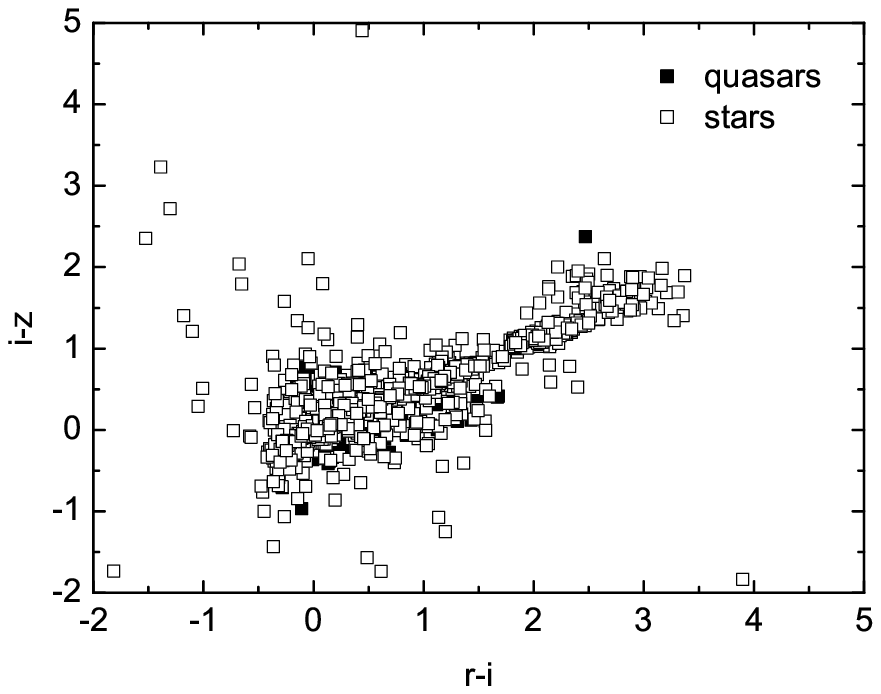}
\includegraphics[bb=0 0 300 240,width=6cm,clip]{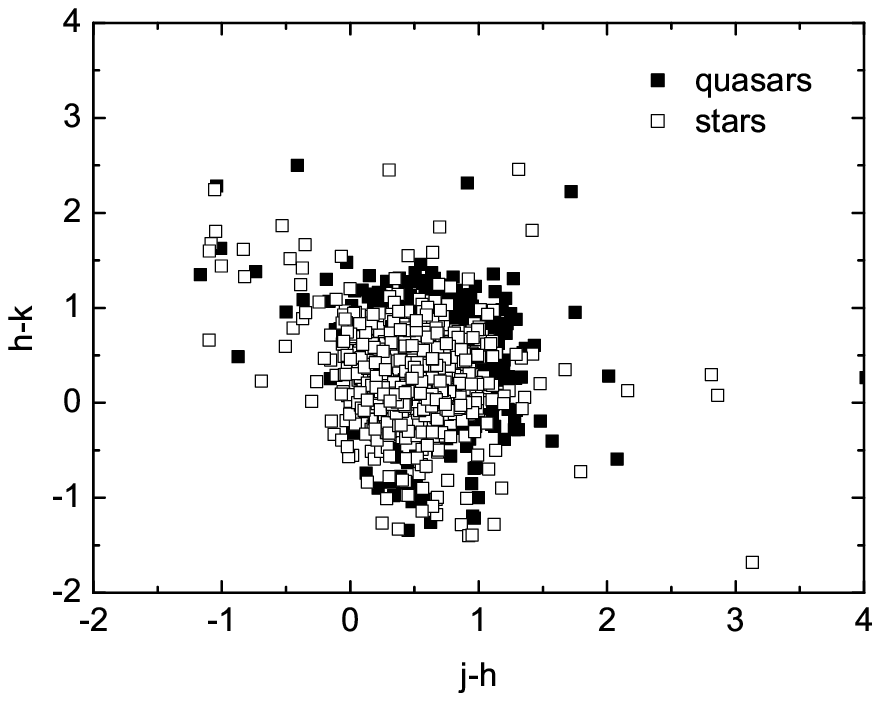}

\includegraphics[bb=0 0 300 240,width=6cm,clip]{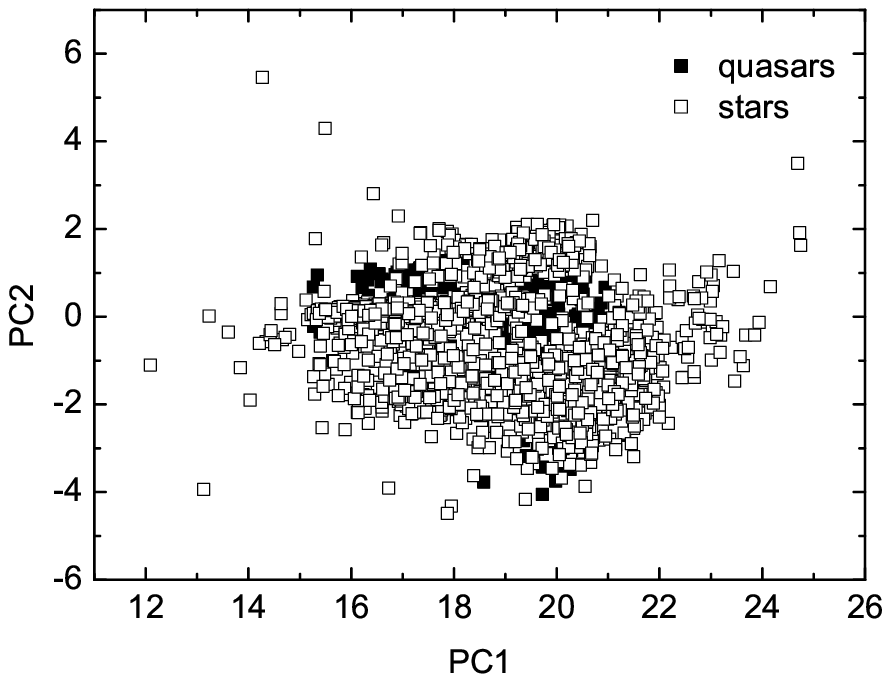}
\includegraphics[bb=0 0 300 240,width=6cm,clip]{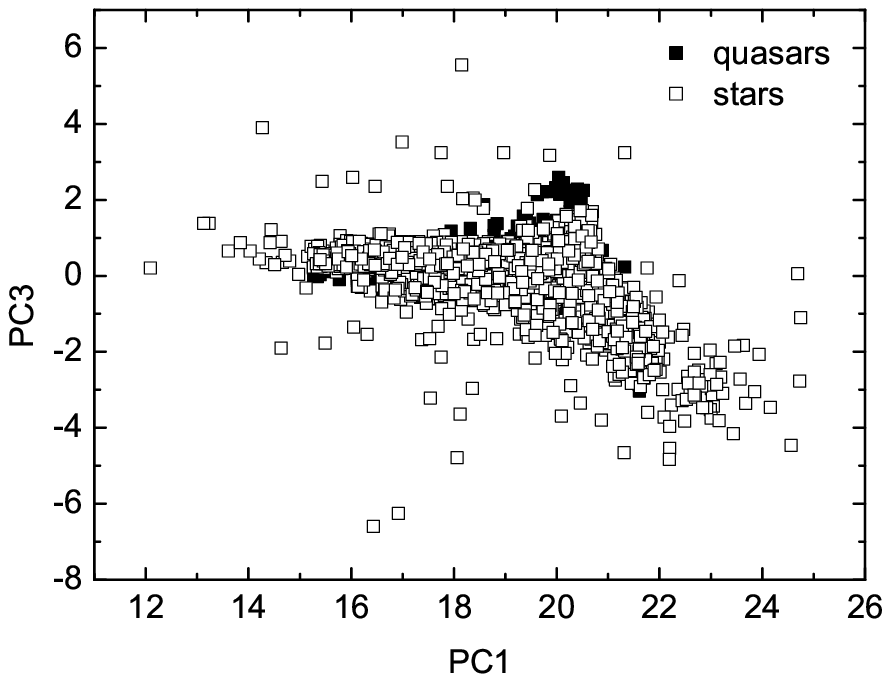}

   \end{center}
   \caption
   { \label{fig:example}
Scatter plots of random subsample (filled squares represent quasars;
open ones represent stars.): the upper four diagrams are color-color
diagrams; the lower two diagrams are PC1 vs. PC2 and PC1 vs. PC3.}
   \end{figure*}

\section{The Classification Algorithms}

\subsection{Kd-tree}

K-dimensional tree (kd-tree), as a computer science term, is a
space-partitioning data structure for organizing points in a
$k$-dimensional space (Bentley, 1975). For more information about
kd-tree, we refer to http://en.wikipedia.org/wiki/Kdtree. A kd-tree
uses only splitting planes that are perpendicular to one of the
coordinate system axes. In addition, in the typical definition every
node of a kd-tree, from the root to the leaves, stores a point. As a
consequence, each splitting plane must go through one of the points
in the kd-tree. Kd-trees are a variant that store data only in leaf
nodes. It is worth noting that in an alternative definition of
kd-tree the points are stored in its leaf nodes only, although each
splitting plane still goes through one of the points. Technically,
the letter $k$ refers to the numbers of dimensions. A 3-dimensional
kd-tree can be called as 3d-tree. A graphical representation of a
3d-tree is shown in Figure 2. Kd-tree organizes a set of datapoints
in k-dimensional space in such a way that once built, whenever a
query arrives requesting a list all points in a neighborhood, the
query can be answered quickly without needing to scan every single
point. Each tree node represents a subvolume of the parameter space,
with the root node containing the entire $k$ dimensional volume
spanned by the data. Non-leaf nodes have two children, obtained by
splitting the widest dimension of the parent's bounding box, the
left child owning those data points that are strictly less than the
splitting value in the splitting dimension, and the right child
owning the remainder of the parent's data points. Kd-tree is usually
constructed top-down, beginning with the full set of points and then
splitting in the center of the widest dimension. This produces two
child nodes, each with a distinct set of points. A kd-tree can be
constructed by repeating the procedure recursively.

\begin{figure*}
   \begin{center}

\includegraphics[bb=0 0 565 519,width=8cm,clip]{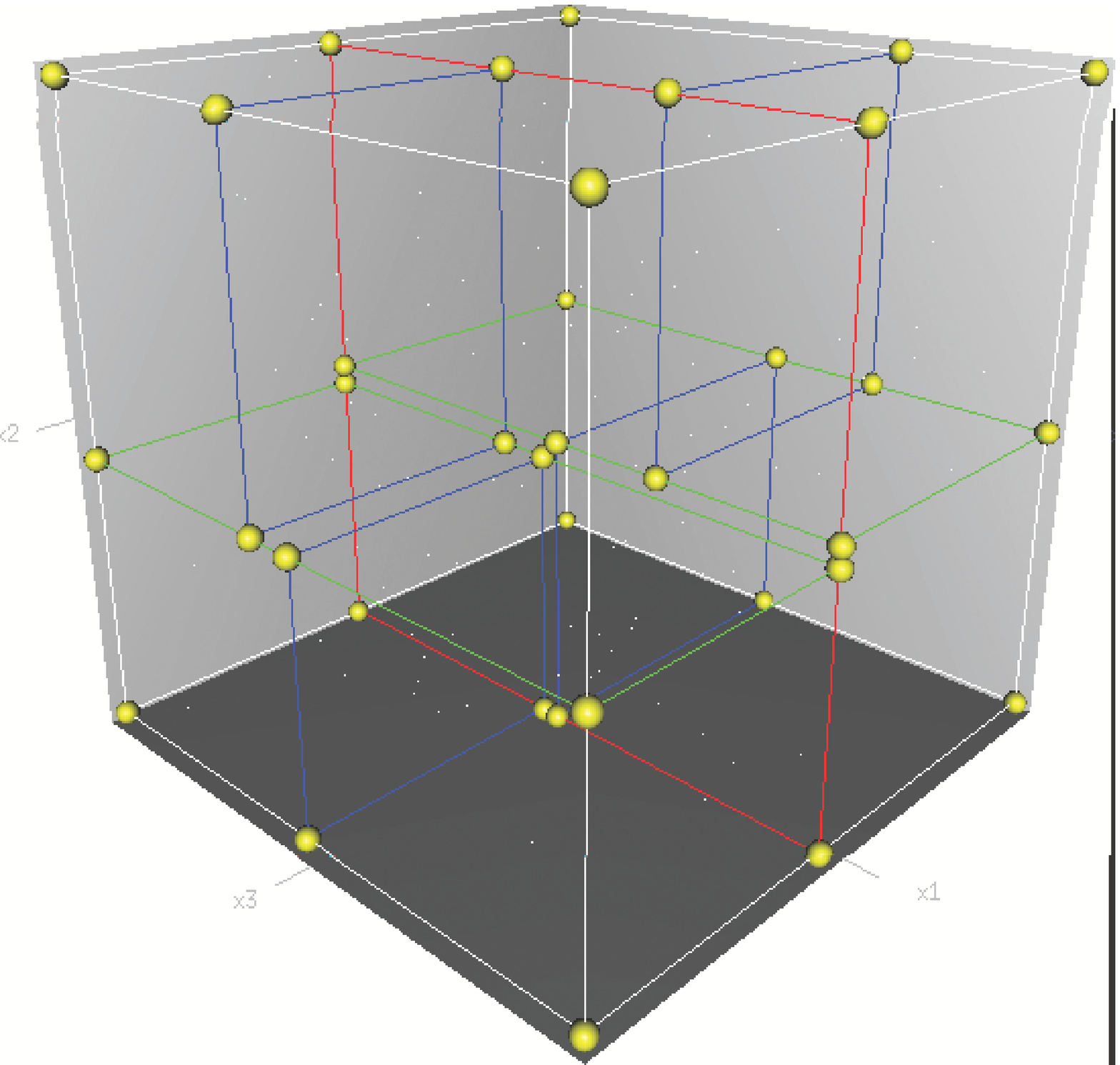}

   \end{center}
   \caption
   { \label{fig:example}
A 3-dimensional kd-tree. The first split (red) cuts the root cell
(white) into two subcells, each of which is then split (green) into
two subcells. Finally, each of those four is split (blue) into two
subcells. Since there is no more splitting, the final eight are
called leaf cells. The yellow spheres represent the tree vertices. }
   \end{figure*}

\subsection{Support Vector Machines}

Support vector machines (SVMs) are a set of related supervised
learning methods used for classification and regression. SVMs can
be considered as a special case of Tikhonov regularization. The
idea of SVMs is to map input vectors nonlinearly into a
high-dimensional feature space and construct the optimal
separating hyperplane in the high-dimensional feature space. SVMs
were originally developed by Vapnik (1995), became popular because
of many attractive features, and promises empirical performance.
SVMs have various parameters that can be tuned for optical
performance, including the kernel function. Popular kernels
consist of linear, polynomial and radial basis function. SVMs also
allow adjusting the soft margin, which is a parameter that
controls the trade-off between smooth and overly complex
functions. Controlling this trade-off is necessary to obtain good
generalization. Functions that represent the training data well
but do not generalize to novel examples are said to have overfit
the data in machine learning terminology. The soft margin is a
tool for SVMs to avoid overfitting (Rohde et~al. 2005).

\subsection{Performance Measurement}

Besides the overall classification accuracy, we use metrics such as
true negative rate, true positive rate, Weighted Accuracy (WA),
G-mean (GM), precision, recall, and F-measure (FM) to evaluate the
performance of classification algorithms (Chen \& Liaw 2004). These
metrics have been widely used for comparison of different
classifiers. All these metrics are functions of the confusion matrix
as shown in Table 3. TP is short for the true positive, FN for the
false negative, FP for the false positive, TN for the true negative.
In the process of classification, quasars are labeled as positive,
stars as negative. The rows of the matrix are actual classes, and
the columns are the predicted classes. Based on Table 3, the
above-mentioned metrics are defined as follows:
\begin{equation}
Accuracy (Acc)= {\frac{TP+TN}{TP+FP+TN+FN}}
\end{equation}
\begin{equation}
True\ Positive\ Rate (Acc^{+}) = {\frac{TP}{TP+FN}} = Recall
\end{equation}
\begin{equation}
True\ Nagative\ Rate (Acc^{-}) = {\frac{TN}{TN+FP}}
\end{equation}
\begin{equation}
Precision = {\frac{TP}{TP+FP}}
\end{equation}
\begin{equation}
F-measure (FM) =
{\frac{2\times{Precision\times{Recall}}}{Precision+Recall}}
\end{equation}
\begin{equation}
G-mean (GM)= (Acc^{-}\times{Acc^{+}})^\frac{1}{2}
\end{equation}
\begin{equation}
Weighted\ Accuracy (WA)=
\beta\times{Acc^{+}}+(1-\beta)\times{Acc^{-}}
\end{equation}

\begin{table*}
 \caption{Confusion matrix. }
 \label{symbols}
 \begin{tabular}{|c|c|c|}
  \hline
   & Predicted Positive Class & Predicted Negative Class
       \\
  \hline
  Actual Positive class & TP (True Positive) & FN (False Negative) \\

  Actual Negative class & FP (False Positive) & TN (True Negative)   \\

  \hline
 \end{tabular}

 \medskip

\end{table*}

Recall is the fraction of actual positive cases that were correct,
and precision is the fraction of the predicted positive cases that
were correctly identified. For any classifier, there is always a
trade off between recall and precision. The Geometric Mean (G-mean)
is useful to determine ``average factors". The F-measure can be
interpreted as a weighted average of the precision and recall.
Weighted Accuracy uses an adjusted parameter $\beta$ to suit
different applications. Here we use equal weights for both true
positive rate and true negative rate; i.e., $\beta$ equals 0.5.
These metrics are commonly used in the information retrieval area as
performance measures.  We will adopt all these measurements to
compare our methods with different patterns. Train-test and ten-fold
cross-validation were carried out to obtain all the performance
metrics.

\section{RESULTS AND DISCUSSION}

Our experiments are performed using the kd-tree java package
(http://www.cs.wlu.edu/~levy/software/kd/) written by Simon D. Levy
and SVMLight which is an implementation of SVMs in C language
(http://svmlight.joachims.org/). The configuration of the PC
computer used was Microsoft Windows XP, Pentium (R) 4, 3.2 GHz CPU,
1.00 GB memory. One advantage of the empirical training set approach
to classification is that additional parameters can be easily
incorporated. More parameters may be taken as inputs. In order to
study which parameters influence the classification accuracy, we
probe different input patterns to separate quasars from stars. We
compare the performance of kd-tree and SVMs with different input
patterns. Our experiment results are shown in Tables 4-8. We
calculate accuracy, true positive rate, true negative rate,
precision, F-measure, G-mean, Weighted Accuracy and running time for
all experiment results. We apply these criteria to determine which
pattern is best. Here we take quasars as the positive class and
stars as the negative one. For Weighted Accuracy, we adopt equal
weights for both true positive rate and true negative rate ($\beta$
equals 0.5).

\subsection{Results of kd-tree}

Firstly, we explore kd-tree to isolate quasars from stars with
different input patterns. Each of the samples is randomly divided
into two parts: two thirds for training a classifier and one third
for testing the classifier to get the classification rate. This
method is usually called train-test method. For different input
patterns, the number of samples is different. The label $Q$ ($Q$ for
quasars) or $S$ ($S$ for stars) as input index is inserted into the
samples and used to build a kd-tree classifier in a supervised way.
Then we use the test samples to get the optimal value of $n$ nearest
neighbors. For each test sample, we need judge if there are more
than half of the $n$ nearest neighbors which are equal to the test
sample's input index to obtain correct or incorrect prediction. So
the $n$ value must be an odd integer to avoid half-and-half case. In
theory, the higher values of $n$ provide smoothing that reduces
vulnerability to noise in the training data. In practical
applications $n$ is typically in units or tens rather than in
hundreds or thousands. We set $n$=11 for this experiment because of
its higher accuracy. The magnitudes in five bands ($u$, $g$, $r$,
$i$, $z$) are taken as the first set of input parameters for
kd-tree, and then the four color index ($u-g$, $g-r$, $r-i$, $i-z$)
and $r$ magnitude are as input patterns. There will be more
information for classification if more parameters are included. $J$,
$H$ and $K_s$ magnitudes (or $J-H$ and $H-K_s$) from 2MASS catalog
are added as extra inputs to build our classifier. We compare the
performance of different input patterns based on PSF magnitudes,
model magnitudes and dereddened magnitudes. The comparison of
different input patterns are listed in Table 4.

Table 4 shows that for any input patterns using kd-tree method, the
accuracy is rather high, more than 94.47\%, with high values of
F-measure, G-mean and Weighted Accuracy, and the running time is
less than 5 minutes. Generally, the performance of similar input
patterns based on model magnitudes adds up to a higher accuracy than
those based on other kinds of magnitudes, those of dereddened
magnitudes are better than those of PSF magnitudes. For these three
kinds of magnitudes, the results based on four colors and $r$
magnitude as input patterns outperform those of the five magnitudes.
The accuracy does not always increase with more features considered,
for example, the accuracy seems to decrease when the input patterns
given parameters $J$, $H$, $K_s$, or $J-H$, $H-K_s$. Only when
appropriate features adopted, the performance is best. In the
situations of fewer features, kd-tree shows better performance and
uses less building time. As shown by Table 4, the four model color
index ($u-g$, $g-r$, $r-i$, $i-z$) and the model $r$ magnitude as
the input pattern obtains the highest accuracy which amounts to
97.26\%, and the highest value of F-measure, G-mean and Weighted
Accuracy which are 96.69\%, 97.14\% and 97.14\%, respectively,
moreover the running time is shorter, not more than 1 minute.

\begin{table*}
 \caption{The comparison of
different input patterns using kd-tree when $n$=11. }
 \label{symbols}
 \begin{tabular}{@{}lcccccccr}
  \hline
  Input patterns & Acc & Acc$^{+}$ & Acc$^{-}$
        & Precision & FM & GM & WA & Time \\
       & (\%)& (\%) &(\%) & (\%)& (\%) & (\%) & (\%) & (s)\\
  \hline
  $u^p,g^p,r^p,i^p,z^p$ & $96.32$ & $95.34$ & $97.02$ & $95.76$ & $95.55$ & $96.17$ & $96.18$ & $27$ \\

  $u^p-g^p,g^p-r^p,r^p-i^p,i^p-z^p,r^p$ & $97.02$ & $96.24$ & $97.58$ & $96.56$ & $96.40$ & $96.90$ & $96.91$ & $48$ \\

  $u^p,g^p,r^p,i^p,z^p,J-H,H-K_s$ & $95.82$ & $94.90$ & $96.48$ & $95.01$ & $94.96$ & $95.69$ & $95.69$ &  $83$ \\
  $u^p-g^p,g^p-r^p,r^p-i^p,i^p-z^p,r^p,J-H,H-K_s$ & $96.62$ & $95.89$ & $97.14$ & $95.96$ & $95.92$ & $96.51$ & $96.52$ &  $166$ \\

  $u^p,g^p,r^p,i^p,z^p,J,H,K_s$ & $94.81$ & $93.91$ & $95.45$ & $93.58$ & $93.75$ & $94.67$ & $94.68$ &  $90$ \\
  $u^p-g^p,g^p-r^p,r^p-i^p,i^p-z^p,r^p,J,H,K_s$ & $95.76$ & $95.08$ & $96.24$ & $94.70$ & $94.89$ & $95.66$ & $95.66$ &  $166$ \\

  $u,g,r,i,z$ & $96.46$ & $95.42$ & $97.20$ & $96.02$ & $95.72$ & $96.31$ & $96.31$ &  $26$ \\
  \textbf{u-g,g-r,r-i,i-z,r} & \textbf{97.26} & $96.41$ & $97.87$ & $96.97$ & \textbf{96.69} & \textbf{97.14} & \textbf{97.14} &  $54$ \\

  $u,g,r,i,z,J-H,H-K_s$ & $95.85$ & $94.90$ & $96.53$ & $95.09$ & $94.99$ & $95.71$ & $95.72$ &  $91$ \\
  $u-g,g-r,r-i,i-z,r,J-H,H-K_s$ & $96.76$ & $96.02$ & $97.28$ & $96.15$ & $96.08$ & $96.65$ & $96.65$ &  $148$ \\

  $u,g,r,i,z,J,H,K_s$ & $94.87$ & $93.88$ & $95.57$ & $93.75$ & $93.81$ & $94.72$ & $94.73$ &  $91$ \\
  $u-g,g-r,r-i,i-z,r,J,H,K_s$ & $95.85$ & $95.09$ & $96.39$ & $94.90$ & $95.00$ & $95.74$ & $95.74$ &  $167$ \\

  $u',g',r',i',z'$ & $96.41$ & $95.42$ & $97.10$ & $95.88$ & $95.65$ & $96.26$ & $96.26$ &  $25$ \\
  $u'-g',g'-r',r'-i',i'-z',r'$ & $97.19$ & $96.37$ & $97.76$ & $96.82$ & $96.60$ & $97.07$ & $97.07$ &  $44$ \\

  $u',g',r',i',z',J-H,H-K_s$ & $95.8$ & $95.00$ & $96.47$ & $95.01$ & $95.00$ & $95.73$ & $95.74$ &  $81$ \\
  $u'-g',g'-r',r'-i',i'-z',r',J-H,H-K_s$ & $96.68$ & $95.97$ & $97.18$ & $96.00$ & $95.99$ & $96.57$ & $96.57$ &  $151$ \\

  $u',g',r',i',z',J,H,K_s$ & $94.73$ & $93.78$ & $95.41$ & $93.52$ & $93.65$ & $94.59$ & $94.59$ &  $89$ \\
  $u'-g',g'-r',r'-i',i'-z',r',J,H,K_s$ & $95.76$ & $95.04$ & $96.27$ & $94.74$ & $94.89$ & $95.65$ & $95.65$ &  $178$ \\

  \hline
 \end{tabular}
 \medskip

\end{table*}

From the above results, we conclude that the four model colors
($u-g$, $g-r$, $r-i$, $i-z$) and the model $r$ magnitude is the best
input pattern for kd-tree when setting $n=11$. Now we adopt such
pattern as input pattern and investigate the influence of the $n$
value on the performance of kd-tree. We change the value of $n$ for
different experiments. By comparing accuracy, F-measure, G-mean and
Weighted Accuracy of classification and the running time taken to
build a classifier, we estimate the efficiency and effectiveness of
the classifiers created by different $n$ values in the experiments.
Through the attempts, we obtain the optimal $n$ value of nearest
neighbors. Here we adopt the odd integer of $n$ from 3 to 29 in our
experiments. Table 5 indicates that the highest accuracy of
classification is 97.263\% when $n$=11, and the next highest results
are 97.262\% and 97.252\% when $n$=7 and $n$=9, respectively. The
running time is longer when the value of $n$ is bigger in our
experiment. As $n$=7, the simultaneous highest values of F-measure,
G-mean and Weighted Accuracy are 96.690\%, 97.149\% and 97.151\%,
respectively. We also see that the true positive rate is higher for
$n$=7 than for $n$=9 or 11, which means that the classifier we build
using $n$=7 gives high prediction accuracy over the quasar class,
while maintaining reasonable accuracy for the star class.

\begin{table*}
\caption{The comparison of different $n$ value for kd-tree with four
model colors ($u-g$, $g-r$, $r-i$, $i-z$) and model $r$ magnitude as
input pattern. }
\begin{tabular}{lcccccccl}
\hline
  $n$ & Acc & Acc$^{+}$ & Acc$^{-}$& Precision & FM & GM & WA & Time \\
       & (\%)& (\%) & (\%)& (\%)&(\%)  & (\%) & (\%) &(s) \\
\hline
  3 & $97.167$ & $96.479$ & $97.654$ & $96.677$ & $96.578$ & $97.065$ & $97.066$ & $29$ \\
  5 & $97.239$ & $96.554$ & $97.723$ & $96.775$ & $96.665$ & $97.137$ & $97.139$ & $35$ \\
  \textbf{7} & \textbf{97.262} & $96.503$ & $97.799$ & $96.877$ & \textbf{96.690} & \textbf{97.149} & \textbf{97.151} & $41$ \\
  9 & $97.252$ & $96.452$ & $97.818$ & $96.902$ & $96.677$ & $97.133$ & $97.135$ & $46$ \\
  \textbf{11} & \textbf{97.263} & $96.413$ & $97.866$ & $96.966$ & $96.689$ & $97.137$ & $97.140$ & $50$ \\
  13 & $97.228$ & $96.412$ & $97.804$ & $96.882$ & $96.646$ & $97.106$ & $97.108$ & $54$ \\
  15 & $97.187$ & $96.342$ & $97.785$ & $96.853$ & $96.597$ & $97.061$ & $97.064$ & $57$ \\
  17 & $97.130$ & $96.227$ & $97.768$ & $96.826$ & $96.526$ & $96.994$ & $96.998$ & $61$ \\
 19 & $97.102$ & $96.153$ & $97.774$ & $96.831$ & $96.491$ & $96.960$ & $96.964$ & $64$ \\
 21 & $97.081$ & $96.149$ & $97.740$ & $96.785$ & $96.466$ & $96.941$ & $96.945$ & $67$ \\
  23 & $97.064$ & $96.113$ & $97.737$ & $96.780$ & $96.445$ & $96.922$ & $96.925$ & $70$ \\
 25 & $97.043$ & $96.082$ & $97.723$ & $96.760$ & $96.420$ & $96.899$ & $96.903$ & $73$ \\
 27 & $97.004$ & $96.023$ & $97.698$ & $96.724$ & $96.372$ & $96.857$ & $96.861$ & $76$ \\
  29 & $96.968$ & $95.956$ & $97.684$ & $96.702$ & $96.328$ & $96.816$ & $96.820$& $78$
  \\
\hline
\end{tabular}
\end{table*}

\subsection{Results of SVMs}

Since the best input pattern is four model colors ($u-g$, $g-r$,
$r-i$, $i-z$) and model $r$ magnitude for kd-tree, we apply such
input pattern to create the SVM classifier. The kernel function of
SVMs we choose is the radial basis function (RBF). When using RBF
SVMs, there are two adjusted parameters: $\gamma$ is the parameter
in RBF kernel and $c$ is the trade-off between training error and
margin. Here we try to compare the classifier created by the
different values of these two parameters in our experiments, and the
results are listed in Table 6. It can be seen that the best accuracy
(97.50\%) is obtained using RBF SVM classifier with $\gamma$=5 and
$c$=1 or 5 and the building time is 21 min and 28 min, respectively.
However, the highest F-measure value is 97.78\% with $\gamma$=8 and
$c$=1, and the highest values of G-mean and Weighted Accuracy are
97.41\% and 97.41\% with $\gamma$=5 and $c$=5. We also find that the
true positive rate when $\gamma$=5 and $c$=5 is superior to that
when $\gamma$=5 and $c$=1, but the former takes more training time.
Table 6 shows that in the situation of the smaller $c$ value, less
running time is generally taken. So when $\gamma$ equals 5 and $c$
equals 0.1, we take the least time to build the SVM classifier,
while $\gamma$ equals 0.01 and $c$ equals 1000, the time taken adds
up to 1 day and 14 hours.

On account of the highest values of G-mean and Weighted Accuracy in
Table 6, the optimal values of $\gamma$ and $c$ are 5 for RBF SVMs.
Then setting $\gamma$=5 and $c$=5, we compute accuracy, true
positive rate, true negative rate, precision, F-measure, G-mean and
Weighted Accuracy for different input patterns using RBF SVMs in
Table 7. Clearly based on accuracy, F-measure, G-mean and Weighted
Accuracy, the input pattern of ($u-g$, $g-r$, $r-i$, $i-z$, $r$) is
the optimal pattern. Using four colors and $r$ magnitude ($u-g$,
$g-r$, $r-i$, $i-z$, $r$) as input pattern, the performance of SVMs
is better than the five magnitudes ($u$, $g$, $r$, $i$, $z$).
Similar to the result of kd-tree, the performance based on the model
magnitudes outperforms that based on the dereddened magnitudes,
which is superior to that based on the PSF magnitudes. When adding
more parameters from near infrared band, the performance doesn't
improve, even decrease. This shows that $J-H$ and $H-K_s$ contribute
little information for classification.

\begin{table*}
 \caption{The comparison of different options using RBF SVMs with four
model colors ($u-g$, $g-r$, $r-i$, $i-z$) and model $r$ magnitude as
input pattern. }
 \label{symbols}
 \begin{tabular}{@{}lccccccccl}
  \hline
  Algorithm &Soft margin & Acc & Acc$^{+}$ & Acc$^{-}$
        & Precision & FM & GM & WA & Time \\
   (RBF kernel)& $c$   &(\%) & (\%) & (\%)&(\%) & (\%) & (\%) & (\%) & (m) \\
  \hline
  $\gamma=5$&$c=0.1$& $96.85$ & $95.14$ & $98.07$ & $97.22$ & $97.64$ & $96.59$ & $96.60$ & $17$ \\

  $\gamma=0.01$&$c=1$& $92.09$ & $88.69$ & $94.50$ & $91.95$ & $93.21$ & $91.55$ & $91.60$ & $32$ \\

  \textbf{$\gamma$=5}&\textbf{c=1} & \textbf{97.50} & $96.67$ & $98.09$ & $97.27$ & $97.68$ & $97.38$ & $97.38$ & $21$ \\

  $\gamma=8$&$c=1$& $97.48$ & $96.50$ & $98.17$ & $97.39$ &  \textbf{97.78} & $97.33$ & $97.34$ & $30$ \\

  $\gamma=0.01$&$c=10$ & $92.90$ & $89.37$ & $95.40$ & $93.22$ & $94.30$ & $92.34$ & $92.39$ & $47$ \\

  $\gamma=5$&$c=10$ & $97.41$ & $96.86$ & $97.79$ & $96.88$ & $97.33$ & $97.32$ & $97.33$ & $69$ \\

  \textbf{$\gamma$=5}&\textbf{c=5} & \textbf{97.50} & $96.88$ & $97.93$ & $97.07$ & $97.50$ & \textbf{97.41} & \textbf{97.41} & $49$ \\

  $\gamma=8$&$c=5$ & $97.38$ & $96.61$ & $97.92$ & $97.04$ & $97.48$ & $97.26$ & $97.26$ & $54$ \\

$\gamma=0.01$&$c=1000$
& $94.55$ & $94.26$ & $96.02$ & $92.47$ & $93.60$ & $94.50$ & $94.50$ & $2280$ \\

  \hline
 \end{tabular}

 \medskip

\end{table*}

\begin{table*}
 \caption{The comparison of different input patterns using RBF SVM when
 $c$=5 and $\gamma$=5. }
 \label{symbols}
 \begin{tabular}{@{}lcccccccl}
  \hline
  Input patterns & Acc & Acc$^{+}$ & Acc$^{-}$
        & Precision & FM & GM & WA & Time \\
       &(\%) & (\%) &(\%) &(\%) & (\%) & (\%) & (\%) & (m)\\
  \hline
  $u^p,g^p,r^p,i^p,z^p$ & $96.97$ & $96.47$ & $97.40$ & $96.97$ & $97.19$ & $96.94$ & $96.94$ & $58$ \\

  $u^p-g^p,g^p-r^p,r^p-i^p,i^p-z^p,r^p$ & $97.39$ & $96.95$ & $97.77$ & $97.39$ & $97.58$ & $97.36$ & $97.36$ & $42$ \\

  $u,g,r,i,z$ & $97.15$ & $96.64$ & $97.59$ & $97.16$ & $97.37$ & $97.11$ & $97.11$ & $61$ \\

  \textbf{u-g,g-r,r-i,i-z,r} & \textbf{97.50} & $96.97$ & $97.93$ & $97.49$ & \textbf{97.71} & \textbf{97.45} & \textbf{97.45} & $44$ \\

 $u-g,g-r,r-i,i-z,r,J-H,H-K_s$ & $97.17$ & $96.16$ & $97.93$ & $97.18$ & $97.55$ & $97.04$ & $97.04$ & $62$ \\

 $u',g',r',i',r'$ & $97.15$ & $96.72$ & $97.53$ & $97.16$ & $97.34$ & $97.12$ & $97.12$ & $60$ \\

 $u'-g',g'-r',r'-i',i'-z',r'$ & $97.47$ & $97.02$ & $97.85$ & $97.47$ & $97.66$ & $97.44$ & $97.44$ & $43$ \\

  \hline
 \end{tabular}

 \medskip

\end{table*}

Finally, we use 10-fold cross-validation to evaluate the performance
and the speed to build the classifiers of kd-tree and SVMs and adopt
the four colors ($u-g$, $g-r$, $r-i$, $i-z$) and $r$ magnitude as
input pattern in order to compare their performance.
Cross-validation is a generally applicable and very useful technique
for many tasks often encountered in machine learning, such as
accuracy estimation, feature selection or parameter tuning.
Cross-validation is used within a wide range of machine learning
approaches, such as kd-tree and SVMs. K-fold cross-validation is an
important cross-validation method applicable for data set with
moderate size. The data is randomly partitioned into $k$ subsamples.
Each time, one of the subsamples is retained as the testing data,
and the remaining $k-1$ subsamples are put together to form the
training data. The cross-validation process is then repeated $k$
times, then the mean error and the evaluated value across all trials
is computed. The comparison of the efficiency and effectiveness of
the two methods is based on the metrics such as true negative rate,
true positive rate, Weighted Accuracy, G-means, precision, recall,
and F-measure to evaluate the performance of learning algorithms.

Given the metrics in Table 8, the best results only with respect
to accuracy is obtained using RBF SVM classifier with $\gamma$=5
and $c$=5, when the accuracy of classification is 97.65\% and the
standard error is 0.22\%. Furthermore, we get the accuracy of
97.65\% when $\gamma$=2 and $c$=10, and 97.64\% when $\gamma$=5
and $c$=1, but the standard error of the latter is the smallest
among these three best cases. The highest F-measure is 98.01\%
with $\gamma$=10 and $c$=1; the highest G-mean and Weighted
Accuracy value are 97.55\% and 97.56\%. Kd-tree obtains the best
result when $n$=9. The highest values of accuracy, F-measure,
G-mean and Weighted Accuracy amounts to 97.45\%, 97.66\%, 97.32\%
and 97.32\%, respectively, and these results are a little better
than those when $n$=11. Clearly based on the metrics in Table 8,
we can hardly tell the difference between kd-tree and SVMs. SVMs
is slightly better than kd-tree in G-mean, F-measure and Weighted
Accuracy. Since the accuracy of the two learning algorithms is
more than 97.0\%, the two methods are effective classifiers to
isolate quasars from stars.

\subsection{Performance Comparison of kd-tree and SVMs}

From the tables above we conclude that kd-tree and SVMs are
comparable to separate quasars from stars in respect of the
accuracy. When only considering the running time, kd-tree is much
faster than SVMs, for the speed of kd-tree is measured by seconds
while that of SVMs is measured by minutes, as shown in Tables~4-7.
Taking into account both accuracy and speed, kd-tree shows its
superiority, because the speed to build the SVM classifier is very
slow. Moreover, the performance obtained by the 10-fold
cross-validation method gets higher accuracy than the train-test
method because the cross-validation method has the advantage of
producing an effectively unbiased error estimate, but it is
computationally expensive (about 10 times longer than train-test
method). As a result, the classifiers trained with kd-tree and SVMs
can be used to classify the unclassified sources and be applicable
to preselect quasar candidates from SDSS and other survey catalogs.

\begin{table*}
 \caption{The comparison of different $n$ or ($\gamma$ and $c$) using 10-fold cross-validate
method with four model colors ($u-g$, $g-r$, $r-i$, $i-z$) and model
$r$ magnitude as input pattern. }
 \label{symbols}
 \begin{tabular}{@{}lccccccll}
  \hline
  $n$/($\gamma,c$) & Acc & Acc$^{+}$ & Acc$^{-}$
        & Precision & FM& GM & WA  \\
       & (\%)& (\%) & (\%)& (\%)& (\%) & (\%) & (\%) \\
  \hline

  $n=13$ & $97.40\pm0.23$ & $96.47\pm0.31$ & $98.06\pm0.34$ & $97.24\pm0.48$ & $97.65\pm0.41$ & $97.26\pm0.23$ & $97.26\pm0.23$ \\

  $n=11$ & $97.42\pm0.25$ & $96.53\pm0.31$ & $98.06\pm0.35$ & $97.24\pm0.48$& $97.65\pm0.41$ & $97.29\pm0.24$ & $97.29\pm0.24$ \\

  \textbf{n=9} & \textbf{97.45$\pm$0.23} & $96.58\pm0.28$ & $98.07\pm0.34$ & $97.26\pm0.47$ & \textbf{97.66$\pm$0.41} & \textbf{97.32$\pm$0.22} & \textbf{97.32$\pm$0.22} \\

   $n=7$ & $97.44\pm0.23$ & $96.60\pm0.30$ & $98.03\pm0.34$ & $97.21\pm0.47$ & $97.62\pm0.41$ & $97.31\pm0.22$ & \textbf{97.32$\pm$0.22} \\
\hline
  \textbf{c=1, $\gamma$=5} & \textbf{97.64$\pm$0.20} & $96.76\pm0.26$ & $98.26\pm0.34$ & $97.52\pm0.47$ & $97.89\pm0.40$ & $97.50\pm0.19$ & $97.51\pm0.19$\\

  $c=1$,$\gamma$=8 & $97.62\pm0.21$ & $96.64\pm0.22$ & $98.31\pm0.33$ & $97.59\pm0.46$ & $97.95\pm0.40$ & $97.47\pm0.19$ & $97.47\pm0.19$ \\

  $c=1$,$\gamma$=10 & $97.59\pm0.22$ & $96.50\pm0.26$ & $98.37\pm0.34$ & $97.66\pm0.48$ & \textbf{98.01$\pm$0.41} & $97.43\pm0.21$ & $97.43\pm0.21$ \\

  $c=1$,$\gamma$=2 & $97.46\pm0.19$ & $96.63\pm0.26$ & $98.04\pm0.32$ & $97.22\pm0.43$ & $97.63\pm0.38$ & $97.33\pm0.17$ & $97.34\pm0.17$  \\

  \textbf{c=5, $\gamma$=5} & \textbf{97.65$\pm$0.22} & $96.97\pm0.24$ & $98.14\pm0.35$ & $97.37\pm0.35$ & $97.75\pm0.42$ & \textbf{97.55$\pm$0.20} & \textbf{97.56$\pm$0.20} \\

  $c=5$,$\gamma$=8 & $97.61\pm0.26$ & $97.40\pm0.28$ & $98.17\pm0.39$ & $97.40\pm0.54$ & $97.78\pm0.46$ & $97.49\pm0.24$ & $97.50\pm0.24$ \\

  $c=5$,$\gamma$=10 & $97.54\pm0.26$ & $96.66\pm0.31$ & $98.17\pm0.37$ & $97.39\pm0.52$ & $97.78\pm0.45$ & $97.41\pm0.25$ & $97.41\pm0.25$ \\

  $c=10$,$\gamma$=5 & $97.61\pm0.26$ & $96.96\pm0.27$ & $98.07\pm0.41$ & $97.27\pm0.56$ & $97.67\pm0.48$ & $97.51\pm0.24$ & $97.51\pm0.24$ \\

  $c=10$,$\gamma$=8 & $97.50\pm0.26$ & $96.74\pm0.30$ & $98.03\pm0.40$ & $97.20\pm0.55$ & $97.61\pm0.48$ & $97.38\pm0.25$ & $97.39\pm0.25$ \\

  $c=10$,$\gamma$=10 & $97.40\pm0.25$ & $96.53\pm0.32$ & $98.02\pm0.36$ & $97.19\pm0.50$ & $97.60\pm0.43$ & $97.27\pm0.24$ & $97.28\pm0.24$ \\

  \textbf{c=10,$\gamma$=2} & \textbf{97.65$\pm$0.22} &  $97.00\pm0.28$  &  $98.10\pm0.35$ & $97.31\pm0.49$ & $97.70\pm0.42$ & \textbf{97.55$\pm$0.20} & $97.55\pm0.20$ \\

  \hline
 \end{tabular}

 \medskip

\end{table*}

The sources inclined to be misclassified due to their intrinsic
properties are equally prone to misclassification wether kd-tree or
SVMs is used. Most of the sources misclassified by kd-tree overlap
those misclassified by SVMs, as proved by the experimental results.
In order to visualize the classification results, we take the
kd-tree method as an example. In Figure~3, we plot the quasars and
misclassified quasars as the function of redshifts. Figure~3 shows
that the peak of the quasar sample lies in the redshift range 1 to
2, while the peak of misclassified quasars lies in the range 2.5 to
4. The highest peak of the redshift distribution for misclassified
quasars occurs at z$\sim$2.8, which is exactly the redshift range in
which the distinction between M stars and quasars becomes
problematic when the Sloan photometric system is used. The
misclassification simply indicates that, no matter what the
classification method is, one is prone to the same biases because of
the very nature of objects. That the peak of misclassified quasars'
$r$-band magnitude is faint is again due to the fact that the
magnitude limit of the spectroscopic sample was fainter for higher
redshift objects. In addition, we investigate the classified result
as the function of magnitude, as shown in Figure~4. From Figure~4,
it is obviously found that the peak of misclassified quasars or
stars (right panel in Figure~3) shifts to the faint magnitude
compared to quasars and stars (left panel in Figure~3). In other
words, the faint sources are inclined to be misclassified, which
possibly results from the small sample size and low S/N ratio for
these faint sources. We further want to know why the misclassified
sources are prone to be misclassified, so we consult the
misclassified quasars and stars from SIMBAD astronomical database
and NASA/IPAC Extragalactic Database (NED). Of the misclassified
stars, the most objects are CV stars, white dwarfs, RR star, carbon
stars, some objects are ultra-violet sources, X-ray sources, radio
sources, blue sources, and HII region, some objects are galaxies and
irregular spirals, a few are quasars. Of 893 misclassified quasars,
most are quasars with faint magnitudes, some are AGN, Seyfert~1,
Seyfert~2, damped Lyman absorbtion and radio sources, a part are 171
unidentified quasars, and the little part are 4 white dwarfs, one CV
star, one AM star and 29 galaxies.

\begin{figure*}
   \begin{center}
\includegraphics[bb=15 19 290 223,width=6cm,clip]{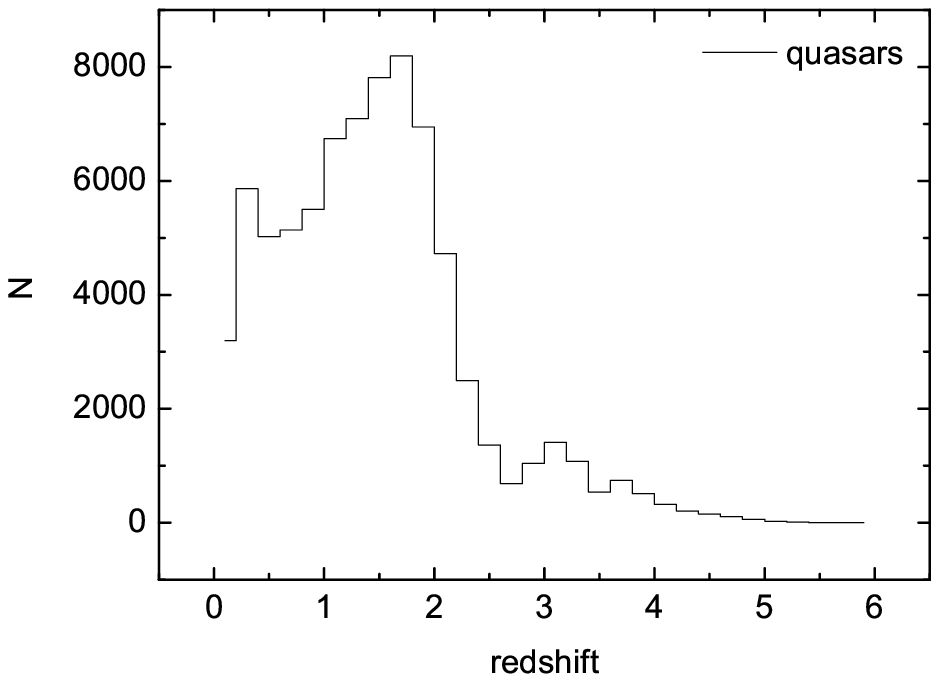}
\includegraphics[bb=15 19 290 223,width=6cm,clip]{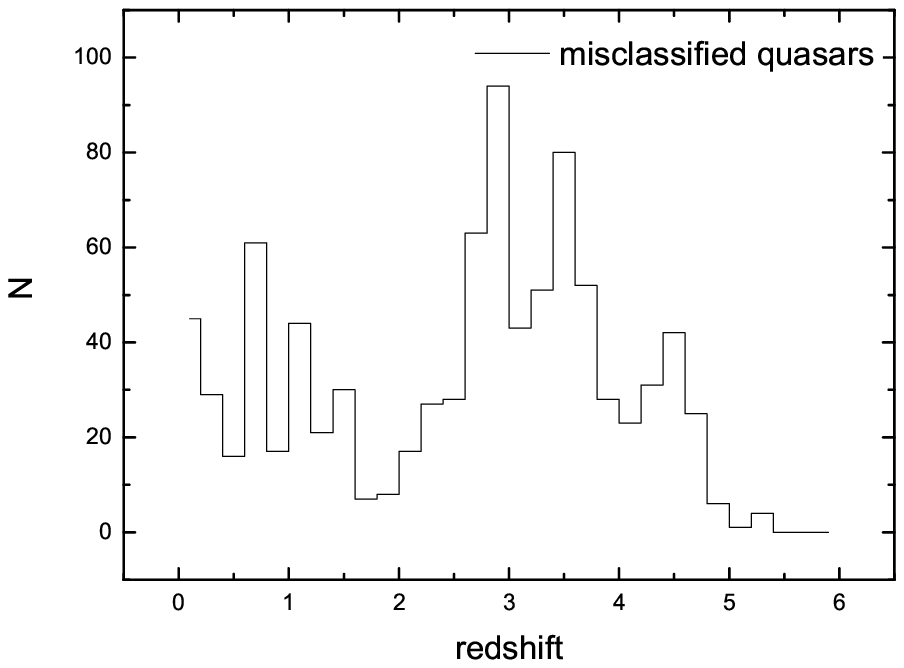}
   \end{center}
   \caption
   { \label{fig:example} The distribution of quasars and
   misclassified quasars as a function of redshift $Z$.
 }
   \end{figure*}

\begin{figure*}
   \begin{center}
\includegraphics[bb=15 19 290 223,width=6cm,clip]{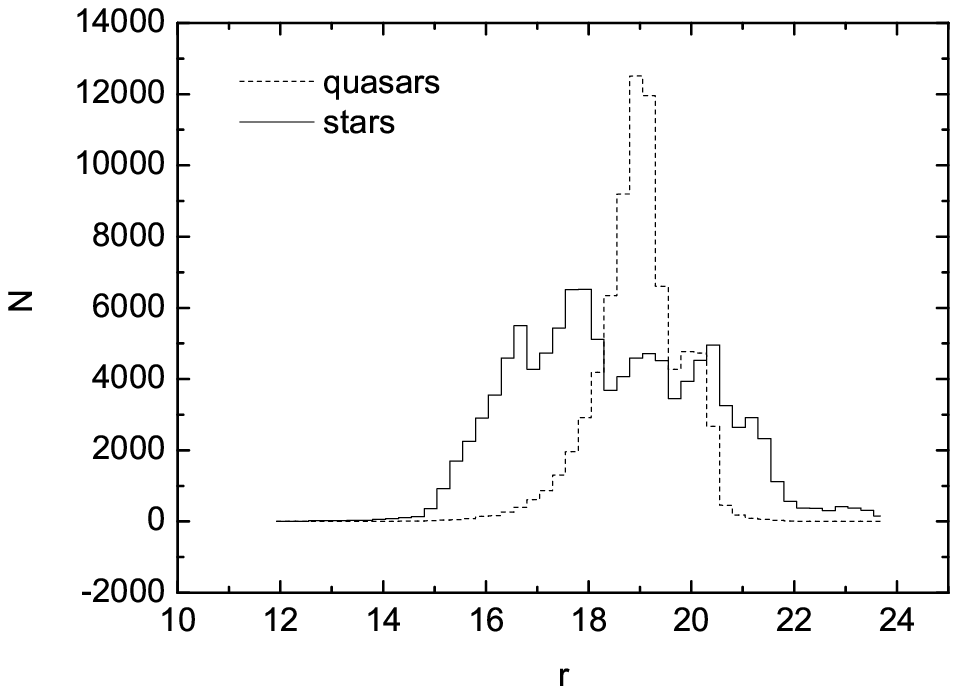}
\includegraphics[bb=15 19 290 223,width=6cm,clip]{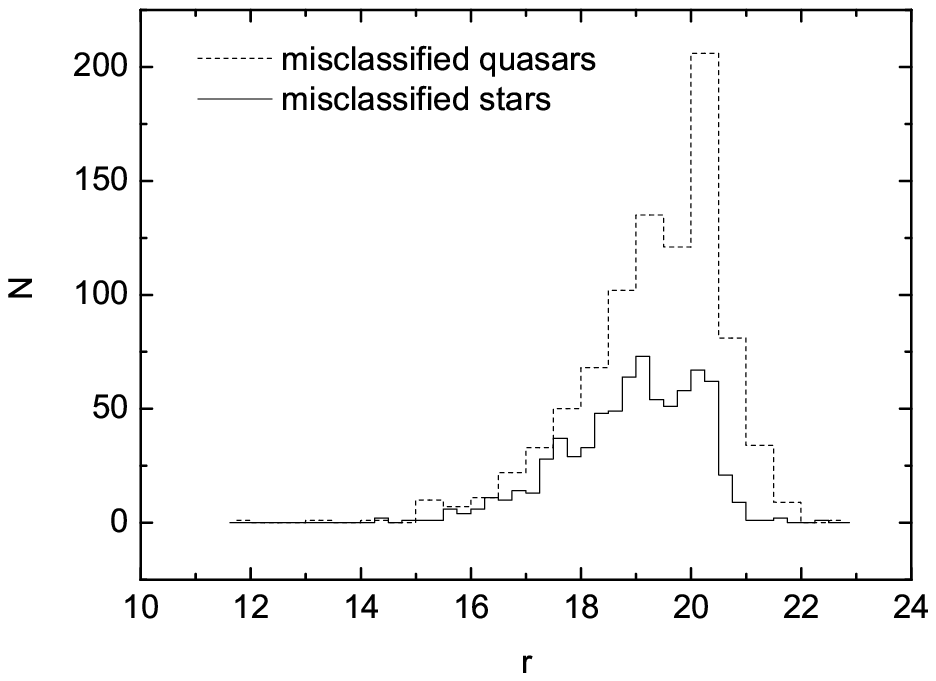}
   \end{center}
   \caption
   { \label{fig:example} The distribution of quasars and
   misclassified quasars (dotted line) as well as stars and misclassified stars (solid line) as a function of model $r$ magnitude.
 }
   \end{figure*}

\section{CONCLUSION}

In this paper we have investigated k-dimensional tree (kd-tree) and
support vector machines (SVMs) applied to the datasets from optical
and infrared band catalogs (SDSS DR5 and 2MASS), and tested it with
different input patterns. We have computed the performance metrics
such as precision and recall, true positive rate and true negative
rate, F-measure, G-mean and Weighted Accuracy to evaluate the
performance of learning algorithms. Based on these metrics from the
results by kd-tree and SVMs, we can not tell clearly which is
superior. Kd-tree and SVMs are comparable to separate quasars from
stars only considering the accuracy. Nevertheless, kd-tree is much
faster to create a classifier than SVMs with respect to the speed.
In real applications, there is one parameter (e.g. the number of
neighbors) to adjust in the kd-tree method while there are two
adjusted parameters (e.g. $\gamma$ and $c$) to control in the SVM
approach when using RBF kernel function. Therefore it is not easy to
modulate optimal parameters and get good performance for SVMs. Given
high accuracy, fast speed and easy modulation of parameters, kd-tree
may be a good choice for classification. Furthermore, both kd-tree
and SVMs show better performance when considering fewer input
parameters. Among the input patterns based on the three kinds of
SDSS magnitudes, the performance of the model magnitudes is the best
and that of the dereddened magnitudes is better than that of the PSF
magnitudes. The input patterns of four colors and $r$ magnitude
($u-g$, $g-r$, $r-i$, $i-z$, $r$) gets better performance than the
five magnitudes ($u$, $g$, $r$, $i$, $z$). We consider more
parameters from 2MASS catalog as extra inputs to our classifiers,
but the results are not better, which is possible attributed to the
bright magnitude limit of $J$, $H$, $K_s$; however other issues
might cause this effect, such as the low measurement precision of
magnitudes. In the experiments, we employ the train-test method and
10-fold cross-validation method to create classifiers. The results
show that the cross-validation method is superior to the train-test
method because the former method avoids the random selection of
sample. When the data are complete or the quality and quantity of
data further improve, the performance of classifiers will improve.
These two approaches can be used to solve the classification
problems faced in astronomy. These classifiers trained by these
methods can be used to classify sources with multi-wavelength
astronomical data and preselect quasar candidates for large surveys,
such as the Chinese Large Sky Area Multi-Object Fiber Spectroscopic
Telescope (LAMOST). Moreover the two methods may be integrated into
the data mining toolkit of Virtual Observatories.

\section{ACKNOWLEDGMENTS}
We are very grateful to referee's constructive and insightful
suggestions as well as helping us improve writing. We would like to
thank LAMOST staff for their help. This paper is funded by National
Natural Science Foundation of China under grant under Grant Nos.
10473013, 90412016 and 10778724. This research has made use of data
products from the SDSS survey and from the Two Micron All Sky Survey
(2MASS). The SDSS is managed by the Astrophysical Research
Consortium for the Participating Institutions. The Participating
Institutions are the American Museum of Natural History,
Astrophysical Institute Potsdam, University of Basel, University of
Cambridge, Case Western Reserve University, University of Chicago,
Drexel University, Fermilab, the Institute for Advanced Study, the
Japan Participation Group, Johns Hopkins University, the Joint
Institute for Nuclear Astrophysics, the Kavli Institute for Particle
Astrophysics and Cosmology, the Korean Scientist Group, the Chinese
Academy of Sciences (LAMOST), Los Alamos National Laboratory, the
Max-Planck-Institute for Astronomy (MPIA), the Max-Planck-Institute
for Astrophysics (MPA), New Mexico State University, Ohio State
University, University of Pittsburgh, University of Portsmouth,
Princeton University, the United States Naval Observatory, and the
University of Washington. 2MASS is a joint project of the University
of Massachusetts and the Infrared Processing and Analysis
Center/California Institute of Technology, funded by the National
Aeronautics and Space Administration and the National Science
Foundation. This research has made use of the NASA/IPAC
Extragalactic Database (NED) which is operated by the Jet Propulsion
Laboratory, California Institute of Technology, under contract with
the National Aeronautics and Space Administration. This research has
also made use of the SIMBAD database, operated at CDS, Strasbourg,
France.

\label{lastpage}

\end{document}